# Performance Comparison Between OpenCV Built in CPU and GPU Functions on Image Processing Operations

Batuhan Hangün*‡, Önder Eyecioğlu**

*Department of Electrical and Electronics Engineering, Nişantaşı University, ISTANBUL

** Department of Computer Engineering, Nişantaşı University, ISTANBUL

(batuhan.hangun, onder.eyecioglu { @nisantasi.edu.tr })

‡ Hangun Batuhan, Department of Electrical and Electronics Engineerinig, Nisantasi University, Istanbul, Turkey, Tel: +90 530 972 20 31,

batuhan.hangun@nisantasi.edu.tr

**Abstract-** Image Processing is a specialized area of Digital Signal Processing which contains various mathematical and algebraic operations such as matrix inversion, transpose of matrix, derivative, convolution, Fourier Transform etc. Operations like those require higher computational capabilities than daily usage purposes of computers. At that point, with increased image sizes and more complex operations, CPUs may be unsatisfactory since they use Serial Processing by default. GPUs are the solution that come up with greater speed compared to CPUs because of their Parallel Processing/Computation nature. A parallel computing platform and programming model named CUDA was created by NVIDIA and implemented by the graphics processing units (GPUs) which were produced by them. In this paper, computing performance of some commonly used Image Processing operations will be compared on OpenCV's built in CPU and GPU functions that use CUDA.

## 1. Introduction

A signal is a function that indicates how a variable change depending on another variable or variables. There are many physical phenomena which may be called as signals such as variation over time in capacitor voltage at an RLC circuit or human voice that attenuate with time or a room temperature at a particular spot etc. [1]. As it was said before, signals are defined mathematically as functions of one or more independent or dependent variables. An audio signal can be represented mathematically by amplitude as a function of time [2]. Concept of DSP is originated at 1960s and 1970s. Since computers were expensive at this period, applications of DSP were limited to some crucial areas. These areas were:

- ➢ Radar and sonar
- ➢ Oil Exploration
- ➢ Medical Imaging
- ➢ Discovery of Space

After the revolution that happened at personal computers between 1980s and 1990s applications of DSP widened and passed the borders of military and government based projects. As a result, DSP entered the public market and the DSP developed with the work of the companies that wanted to profit from this new concept, and it started to spread at a great speed [3]. Since images are two-dimensional signals and these signals need to be processed for various applications, a new subcategory of Signal Processing called Image Processing was born.

During this development of the DSP, the cameras also developed and the image quality and size increased accordingly hence processing of images started to need more computational power and system speed. Because of its nature, images' parallel processing was considered as an efficient and fast way.

Parallel computing using CUDA was applied in many works. Chang et al estimated tensor diffusion using CUDA and MPI. They showed that the processing time of DTI data especially for the data sets with high spatial and temporal resolution was reduced by GPU implementation Framework that was proposed [4]. Saha et al implemented some image enhancement algorithms like darkening filter, negative filter and RGB to Grayscale filter and Recursive Ray Tracing using CUDA [5]. Xin et al successfully accelerated Separable convolution template (SCT) background prediction with CUDA for infrared small target detection [6]. Also Park et al designed and evaluated performances of image processing algorithms on GPUs [7].

In this work, some common used image processing operations such as resizing, thresholding, histogram equalization, edge detection were both run at CPU and GPU, performance criteria for this paper are Time against image size.

## 2. Digital Image and Image Processing

### 2.1 Digital Image

After the images were acquired by sensors, they are reconstructed digitally at memory of devices and it is stored as matrices with MxNxD size. As it was shown in Figure 1.1 intersections of rows and columns named "Pixels (Pixel Elements)". At this representation, M and N means Width and Height. D means Dimension which defines how many values each pixel stores.

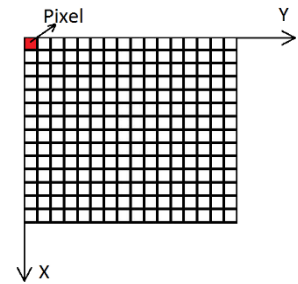

**Fig. 1.** Digital Image Representation

Each pixel can take on a range of values depending on the type of image. For example, at grayscale images, a pixel can only have values between 0(full black) – 255(full white). As it was mentioned before, images may be defined as two-dimensional functions, $f(x,y)$, where x and y are spatial coordinates those represents row number and column number respectively. Value of $f(x,y)$ or amplitude at $f$ at any given (x, y) coordinates is called as "intensity" or "gray level" of image at given coordinates. For any $f$, x, y, and intensity values are limited and discrete quantized images are called **Digital Images** [8].

### 2.2 Image Processing

Image Processing, a sub-field of Signal Processing, is the processing of digital images in computers through algorithms. Because of its nature, it deals with operations of matrices which contain pixels. This is where we need advantages of Parallel Processing since bigger matrix sizes cause greater computation times. The next subtitle will describe the image processing operations used to test GPU performance.

### 2.2.1 Image Resizing

For an image with M x N size, M and N means respectively "Length" and "Width" of digital image which defines how big does an image looks on the screen. Simply, **Image Resizing** is an operation of adjusting M and N. Various interpolation methods like "Bilinear Interpolation", "Bicubic Interpolation", "Nearest-neighbor Interpolation" can be used to resize images. At this paper, "Bilinear Interpolation" is used to resize images.

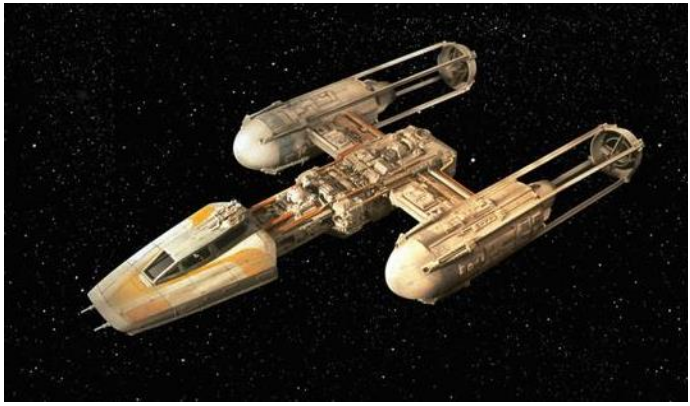

**Fig. 2.** Original Image

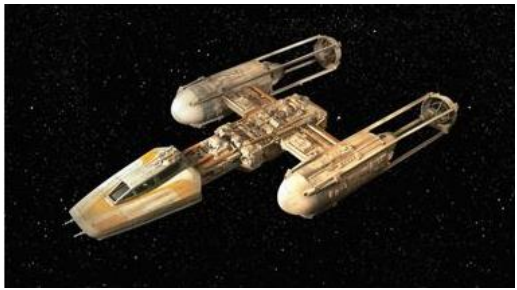

**Fig. 3.** Original image resized to %50 of original image

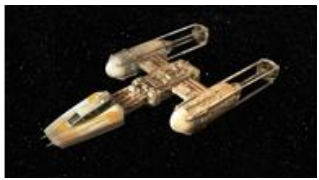

**Fig. 4.** Original image resized to %25 of original image

*2.2.2 Thresholding*

In some applications it may be desirable to binaryize digital images. Operation to create binary images is named **Thresholding** and the k value, which determines which intensity values are 1 and which intensity values are 0, is also called the **threshold**.

$$s = T(r) = \begin{cases} 0, & r < k \\ 1, & r \geq k \end{cases}$$

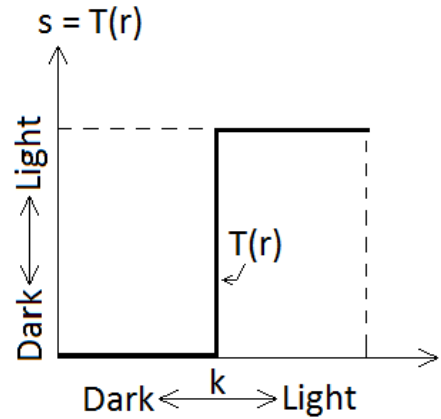

**Fig. 5.** Thresholding function

At this paper, Otsu's method [9] was used for thresholding operations.

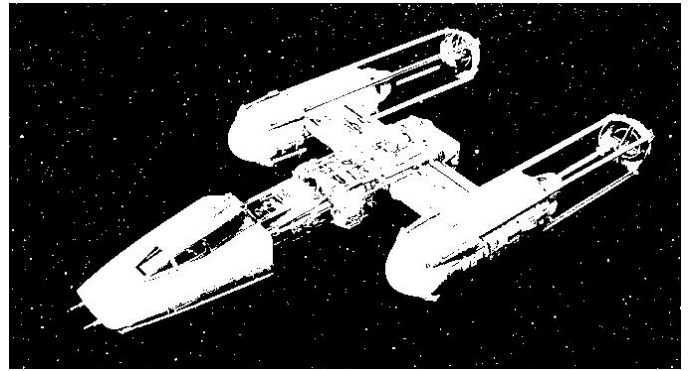

**Fig. 6.** Thresholded image by using Otsu's method

*2.2.3 Histogram Equalization*

**Histogram** is the graphical representation of intensity distribution of an image. At some cases, images with bad contrast may be obtained due to enviroment and image sensor. **Histogram Equalization** is an operation that enhances images with bad contrast. Basically Histogram Equalization is about mapping one image distribution to another one which is wider and uniform distribution of intensity values).

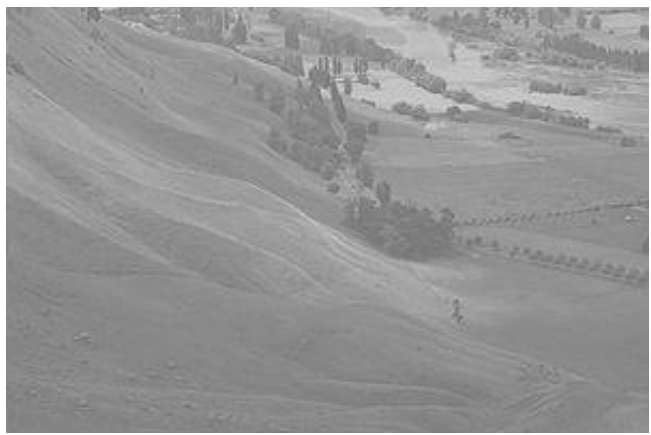

**Fig. 7.** Image with bad contrast

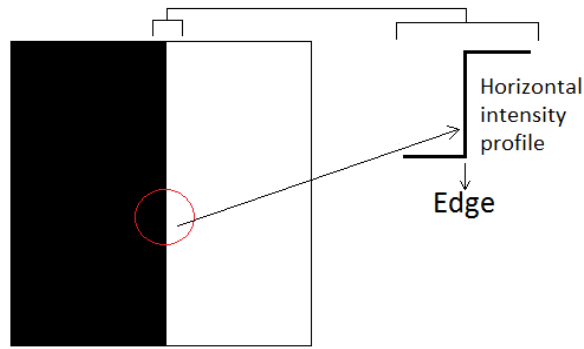

**Fig. 9.** Detecting of edge in an image

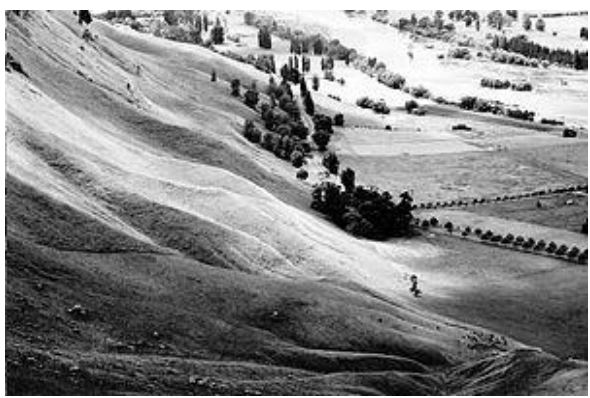

**Fig. 8.** Image contrast enhanced with Histogram Equalization

*2.2.4 Edge Detection*

Edges are the locations where image intensities changed sharply. Edge Detection is a method which is determining where does this intensity changes occur. At some applications like lane detection, image segmentation etc. Edge Detection is crucial.

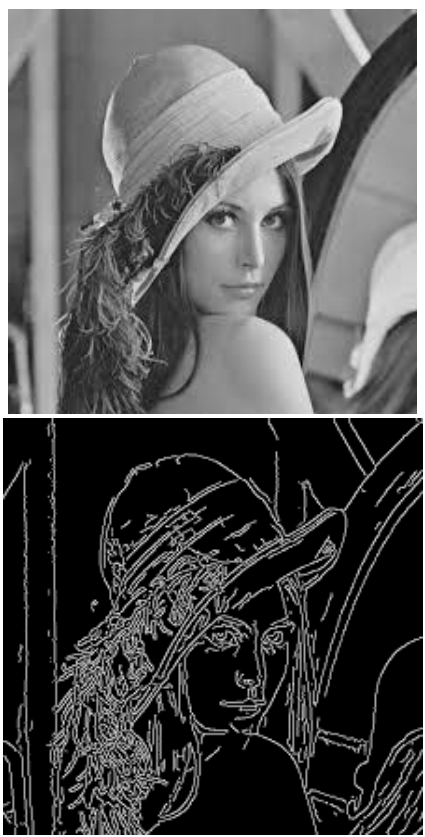

**Fig. 10.** Edge Detection with Canny Edges.

At this paper, Canny Edge Detection [10] was used to detect edges.

**2.3 Image Processing Libraries and Toolkits**

*2.3.1 OpenCV*

OpenCV -synonym for Open Computer Vision- is an open source library which supports Computer Vision and Artificial Intelligence applications development. Gary Bradski who is worked at Intel, launched project OpenCV at 1999. It was written in C/C++ and can run under Windows, Linux and Mac OS X. Interfaces for Java, MATLAB, Python and other languages are also on development [11].

Since it was designed for computational efficiency and focus on real time applications OpenCV is first written in C then updated to and optimized in C ++ and it has advantages of multicore processor. Providing an easy-to-use computer vision library that helps people build computer vision applications quickly is OpenCV's aim. The OpenCV library has more than five hundred functions that cover many areas of Computer Vision [12]. Due to the advantages mentioned above, OpenCV is one of the first options when it comes to Computer Vision applications.

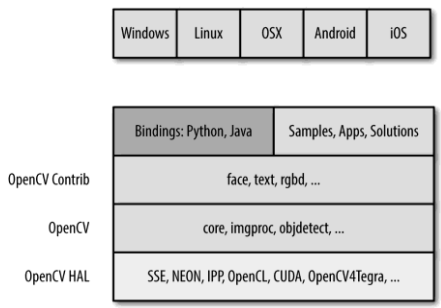

**Fig. 11.** OpenCV Block Diagram with supported Operating Systems [13]

*2.3.2 Compute Unified Device Architecture (CUDA)*

CUDA is an API (application programming interface) and a parallel computing platform model that was created by Nvidia at 2007 [14]. After that day it gained popularity amongst people who needs high computing power with parallelism.  This platform is a software layer which provides direct Access to GPU's virtual instruction set and parallel computing elements, for the execution of computer kernels.

CUDA can work with programming languages such as C, C++ and Fortran which makes it easier to develop programs. A simple CUDA processing follows the flow that is similar to shown below:

- Data is copied from main memory to GPU memory
- CPU instructs the process to GPU
- GPU execute parallel in each core
- Result is copied from GPU memory to CPU memory

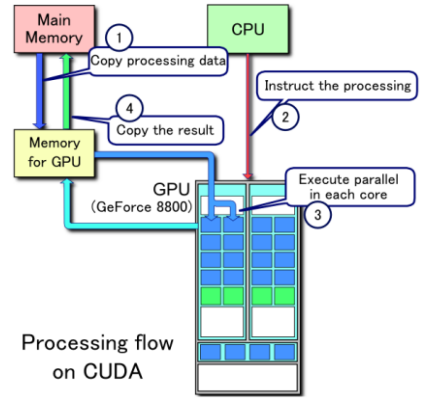

**Fig. 12.** CUDA Processing Flow [15]

**3. Results**

All codes written in C++ using OpenCV 's built-in CPU and GPU functions. System information about PC that runs codes as is follows:

- Operating System: Windows 7 Ultimate 64-bit (6.1, Build 7601) Service Pack 1
- Processor: Intel(R) Core(TM) i7-6700 CPU @ 3.40GHz (8 CPUs), ~3.4GHz
- Memory: 16384MB RAM
- DirectX Version: DirectX 11
- GPU: NVIDIA GeForce GTX 970

At this paper, image size (total number of pixels) vs. time were chosen as performance comparison criterium. Library named "chrono" which is a built-in C++ library was used to measure time.

For resizing and thresholding operations image shown below was used.

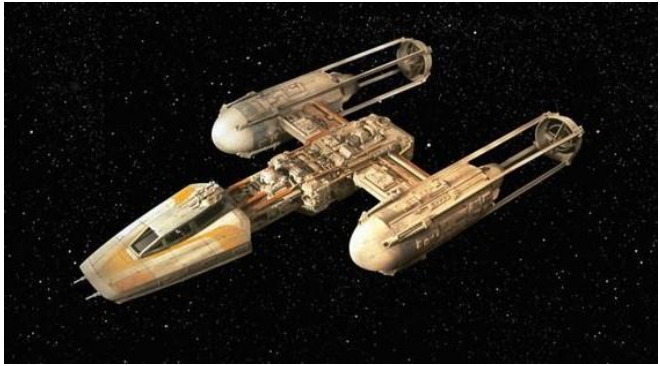

**Fig. 13.** Test image ywing.jpg. Its original size is 564x314. Image was processed for sizes 282x157 846x471, 1128x628, 1410x785, 692x942, 1974x1099, 2256x1256, 2538x1413, 2820x1570, 3102x1727, 3384x1884, 666x2041, 948x2198, 4230x2355, 4512x2512, 4794x2669, 5076x2826, 5358x2983

For histogram equalization process image shown below was used.

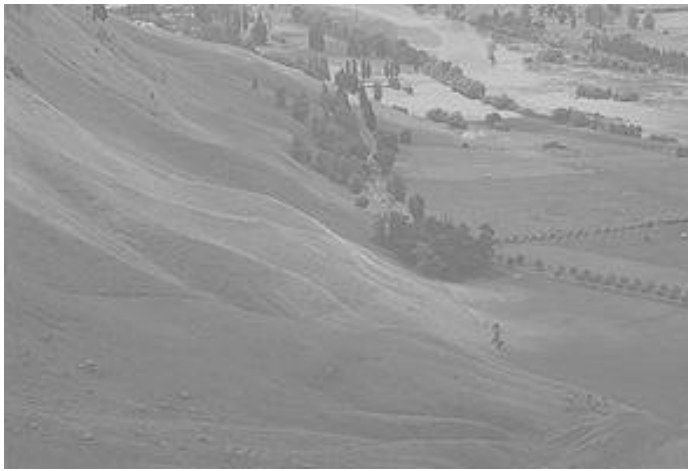

**Fig. 14.** Test image scene_badcontrast.jpg. Its original size is 300x200. Image was processed for sizes 150x100, 450x300, 600x400, 750x500, 900x600, 1050x700, 1200x800, 1350x900, 1500x1000, 1650x1100, 1800x1200, 1950x1300, 2100x1400, 2250x1500, 2400x1600, 2550x1700, 2700x1800, 2850x1900

For edge detection process image shown below was used.

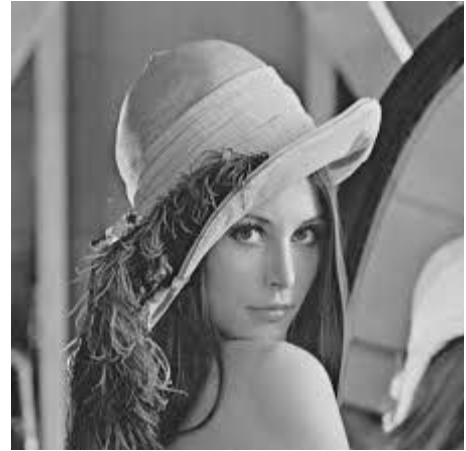

**Fig. 15.** Test image lena.jpg. Its original size is 225x225. Image was processed for sizes 112x112, 338x338, 450x450, 562x562, 675x675, 788x788, 900x900, 1012x1012, 1125x1125, 1238x1238, 1350x1350, 1462x1462, 1575x1575, 1688x1688, 1800x1800, 1912x1912, 2025x2025, 2138x2138

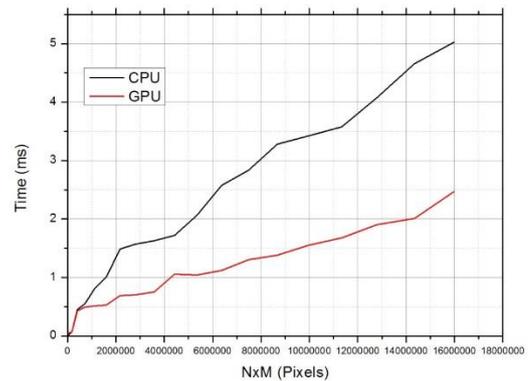

**Fig. 16.** The figure given above shows the comparison of built-in CPU and GPU functions that resize the image in terms of NxM(total number of pixels) and T(time spent in milliseconds). While resizing images, Linear Interpolation method was used.

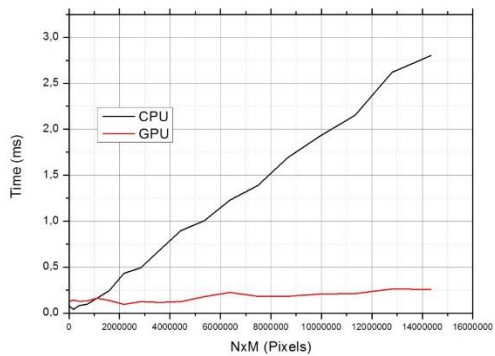

**Fig. 17.** The figure given above shows the comparison of built-in CPU and GPU functions that thresholds the image. While thresholding images, Otsu's method was used.

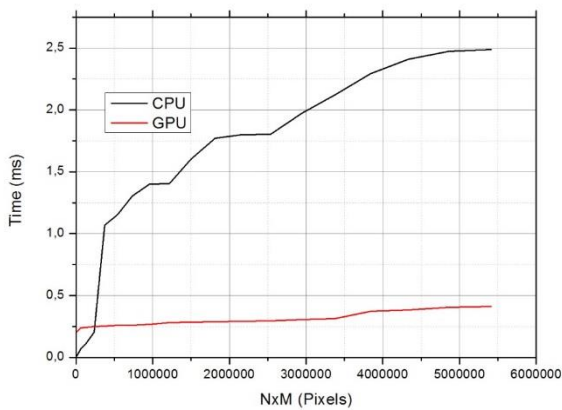

**Fig. 18.** The figure given above shows the comparison of built-in CPU and GPU functions that equalize the histogram of image.

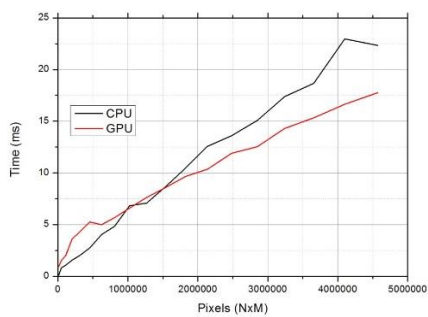

**Fig. 19.** The figures given above shows the comparison of built-in CPU and GPU functions that detects the edges at images. While detecting edges Canny Edge Detector was used.

## 4. Conclusion

In this paper, we measured the time spent at some common used image processing operations with using OpenCV's built-in CPU and OpenCV's built-in GPU functions which are written with CUDA support. Measurements shown that, GPU functions provide a performance improvement because they run in parallel but effects of GPU appear especially when image size increases. As a result of their studies Chang et al [4], Saha et al [5], Xin et al [6] and Park et al [7] proved using GPU and parallelism significantly reduces run time of image processing applications. Studies of Che et al [16], Galizia et al [17] and Yang et al [18] also mentions that they benefitted from performance improvement of GPU.

For future works, instead of OpenCV's built-in GPU functions we plan to use CUDA Toolkit's native functions and libraries to do image processing related operations and evaluate their performances.